\documentclass[aps,prl,amssymb,twocolumn,superscriptaddress,showpacs]{revtex4}

\usepackage{graphicx}
\usepackage{dcolumn}
\usepackage{bm}
\usepackage{color}
\usepackage{epstopdf}

\begin{document}

\newcommand{\beq}{\begin{equation}}
\newcommand{\eeq}{\end{equation}}
\newcommand{\barr}{\begin{eqnarray}}
\newcommand{\earr}{\end{eqnarray}}

\def\bra#1{\langle #1 |}
\def\ket#1{| #1 \rangle}
\def\sinc{\mathop{\text{sinc}}\nolimits}
\def\cV{\mathcal{V}}
\def\cH{\mathcal{H}}
\def\cT{\mathcal{T}}
\renewcommand{\Re}{\mathop{\text{Re}}\nolimits}
\newcommand{\tr}{\mathop{\text{Tr}}\nolimits}

\newcommand{\REV}[1]{\textbf{\color{red}#1}}
\newcommand{\BLUE}[1]{\textbf{\color{blue}#1}}
\newcommand{\GREEN}[1]{\textbf{\color{green}#1}}

\title{Nonlocality and quantum mechanics}

\author{Saverio Pascazio}
\affiliation{Dipartimento di Fisica and MECENAS, Universit\`a di Bari, I-70126 Bari, Italy \\ INFN, Sezione di Bari, I-70126 Bari, Italy}

\date{\today}

\begin{abstract}
A class of nonlocal hidden variable theories is shown to be incompatible with quantum mechanics.
\end{abstract}


\maketitle

Entanglement is one of the most characteristic and puzzling traits of quantum mechanics. Einstein \cite{EPR} and Schr\"odinger \cite{Schr} were the first to point out its counterintuitive aspects (and always nurtured doubts about it). 
Their seminal observations were bounded to the domain of interpretation and 
one had to wait 30 years until Bell \cite{Bell} showed that it is impossible to reproduce the effects of entanglement in terms of classical correlations.
Bell's theorem brought to light a very interesting aspect of the quantum mechanical correlations: they are stronger than their classical counterpart and have a nonlocal flavor. Both these features can be framed in the form of an inequality that is violated by two entangled quantum mechanical particles. When the particles are at a distance, no explanation in terms of a \emph{local} (hidden-variable) model is possible.

Nowadays, entanglement is viewed as a fundamental resource in quantum applications and quantum information \cite{entanglementrev,h4}. At the same time, it preserves its spell, probably because of its mysterious and counterintuitive facets. 
One of these is certainly its role in the subtle interplay between nonlocality and a realistic description of natural phenomena.

The present Letter deals with hidden variables and elaborates on Bell's seminal idea that quantum mechanics is incompatible with their existence. 
Bell considered local hidden variable theories. \emph{Nonlocal} hidden variables were investigated in the 70's \cite{Flato,GS} and have recently attracted renewed attention due to a proposal by Leggett \cite{LeggettFPh}, who considered a particular class of nonlocal hidden-variable theories, that hinge upon rather natural assumptions, yield predictions at variance with quantum mechanics and have the important physical feature of being experimentally falsifiable. These theories were experimentally ruled out a few years after their proposal \cite{groblacher07,Branciard}.

In this Letter we consider a class of theories that are nonlocal, but not ``fully" nonlocal, in the sense that four fifths of the parties involved in an experiment can share nonlocal information among themselves, but not with the fifth one. We show that quantum entanglement yields correlations that are in contradiction even with such a form of nonlocality.

\begin{figure}
\includegraphics[width=0.4\textwidth]{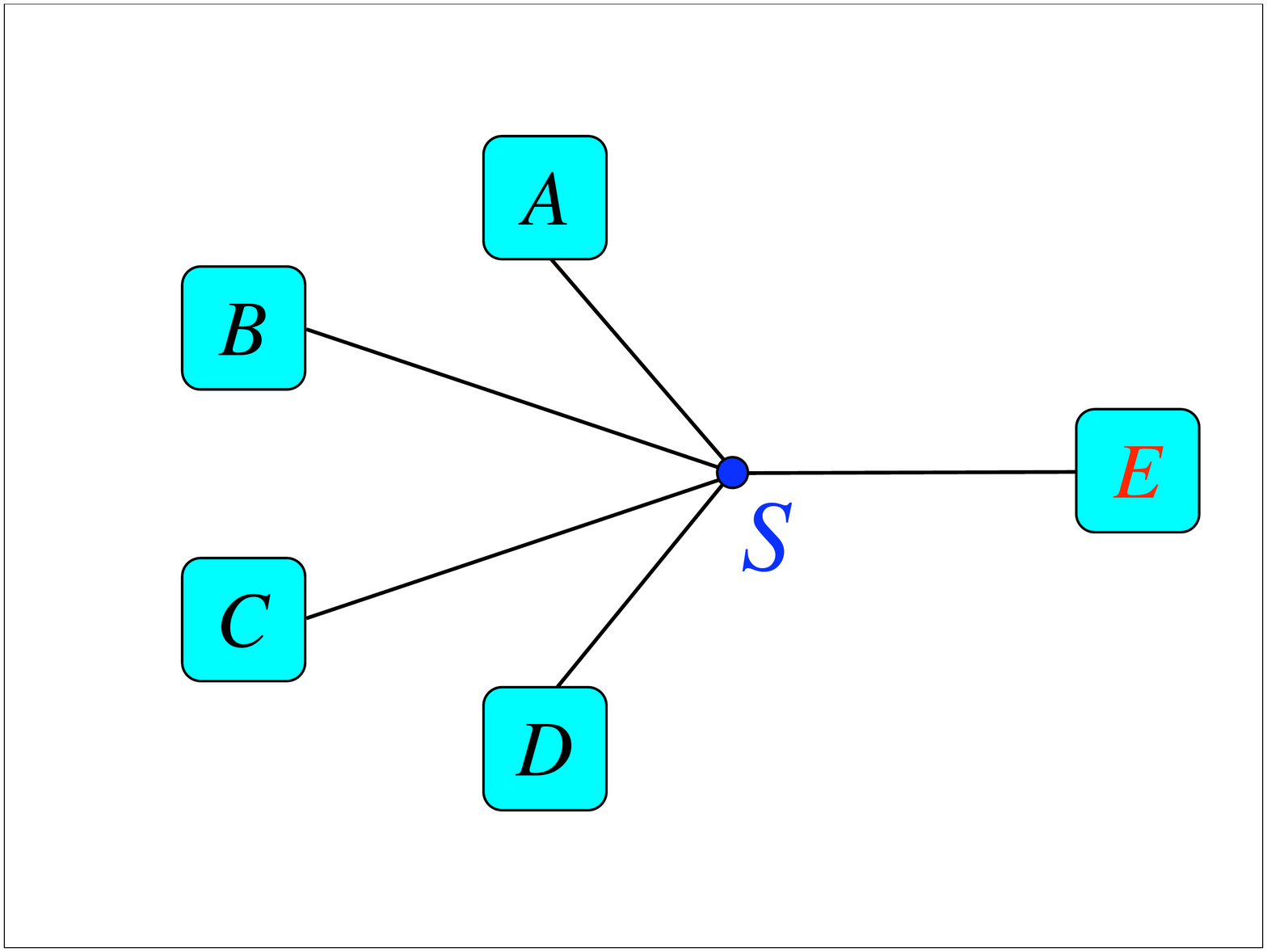}
\caption{Alice, Bob, Charlie, Diana and Esther share five qubits in state (\ref{eq:psi5}) and perform independent (space-like separated) measurements.
For the sake of argument, $A, B, C$ and $D$ are displayed at the left of the particle source $S$, while $E$ is at the right. (Esther means star in Persian, so it is not improper to consider her separated from the others.)
}
\label{fig:abcde}
\end{figure}

Let five parties, Alice, Bob, Charlie, Diana and Esther ($A,B,C,D$ and $E$, respectively), share five qubits in the state
\begin{eqnarray}
|\psi_5\rangle &=& \frac{1}{\sqrt{32}} [ |00000\rangle +
|00001\rangle + |00010\rangle + |00011\rangle  \nonumber \\
 & & + \ |00100\rangle -|00101\rangle - |00110\rangle + |00111\rangle
\nonumber \\
 & & + \ |01000\rangle - |01001\rangle - |01010\rangle + |01011\rangle
 \nonumber \\
 & & + \ |01100\rangle + |01101\rangle + |01110\rangle + |01111\rangle
  \nonumber \\
 & & + \ |10000\rangle +  |10001\rangle - |10010\rangle  - |10011\rangle
  \nonumber \\
 & & + \ |10100\rangle - |10101\rangle + |10110\rangle - |10111\rangle
 \nonumber \\
 & & - \ |11000\rangle + |11001\rangle - |11010\rangle + |11011\rangle
 \nonumber \\
 & & - \ |11100\rangle - |11101\rangle + |11110\rangle + |11111\rangle
 ],
\label{eq:psi5}
\end{eqnarray}
where 0 and 1 denote the two eigenstates of the third Pauli matrix and the qubits belong to $A, B, C, D$ and $E$, respectively. 
State (\ref{eq:psi5}) is very entangled: it is a so-called maximally multipartite entangled state (MMES) \cite{MMES}.
We shall assume henceforth that the measurements done by  $A, B, C, D$ and $E$ are independent (space-like separated). See Fig.\ \ref{fig:abcde}.

Direct calculation yields
\barr
\label{eq:35a}
& & \langle Z_A X_B X_C \openone_D \openone_E\rangle = 1, \\
& & \langle X_A Z_B \openone_C Z_D \openone_E\rangle = 1, \\
& & \langle Y_A Y_B \openone_C \openone_D Z_E \rangle = 1, \\
& & \langle Y_A \openone_B Z_C Y_D  \openone_E \rangle  = 1, \\
& & \langle X_A \openone_B Y_C \openone_D Y_E \rangle = 1, \\
& & \langle Z_A \openone_B \openone_C X_D  X_E \rangle = 1, \\
& & \langle \openone_A Y_B Y_C  X_D \openone_E \rangle = -1, \\
& & \langle \openone_A Z_B Z_C  \openone_D X_E \rangle = 1, \\
& & \langle \openone_A X_B \openone_C Y_D Y_E \rangle = -1, \\
& & \langle \openone_A \openone_B X_C Z_D Z_E \rangle = 1, \\ 
& & \langle X_A X_B Z_C X_D Z_E \rangle = -1, \\
& & \langle X_A Y_B X_C Y_D X_E \rangle = -1, \\
& & \langle Y_A X_B Y_CZ_D X_E \rangle = 1, \\
& & \langle Y_A Z_B X_C X_D Y_E \rangle = 1, \\
& & \langle Z_A Y_B Z_C Z_D Y_E \rangle = 1, \\
& & \langle Z_A Z_B Y_C Y_D Z_E \rangle = 1, 
\label{eq:35}
\earr
where $X, Y, Z$ are Pauli matrices, $\openone$ is the unit operator, the tensor product is omitted and the
average is taken over state (\ref{eq:psi5}). 

It is very simple to see that the above correlations cannot be obtained from a local hidden variable model: indeed, if every quantum mechanical operator is replaced by a (local) classical dichotomic variable, the system (\ref{eq:35a})-(\ref{eq:35}) admits no solution. This comes as no surprise, by virtue of the large entanglement of the MMES (\ref{eq:psi5}).
We now ask whether a \emph{nonlocal} hidden variable description is possible. We start by observing that if a ``full" nonlocality is assumed, the problem is trivial: again, if every operator is replaced by a dichotomic variable, any equation in  (\ref{eq:35a})-(\ref{eq:35}) can be solved independently of any other equation (because the equations are independent and the dichotomic variables, being nonlocal, can depend on the values of the other variables in the same equation). Let us therefore slightly relax our assumptions and require that the outcomes of the measurements of $A, B, C$ and $D$ may \emph{nonlocally} depend on each other, but \emph{not} on those of $E$.
Therefore, Esther's decision to measure, say, spin along $x$, rather than $y$ or $z$, can neither influence the decisions of $A, B, C$ and $D$, nor determine the outcomes of their measurements (except, of course, via local variables, such as e.g.\ those originating in the common source $S$). Moreover, Esther's choices and outcomes are independent of those of  $A, B, C$ and $D$.

At the mathematical level, we assume that
\barr
& & x_A(a,b,c,d,\lambda), \quad  y_A(a,b,c,d,\lambda), \quad
z_A(a,b,c,d,\lambda), \nonumber \\ 
& & \qquad \textrm{and similarly for $B,C$ and $D$,} \label{eq:nonloc} \\ 
& & x_E(e,\lambda), \quad  y_E(e,\lambda), \quad
z_E(e,\lambda),
\label{eq:vardef}
\earr
where $x_A$ is the dichotomic variable \footnote{Strictly speaking, this variable shoud be distinguished from the outcome of a measurement performed along direction $X_A$ \cite{LeggettFPh}, but here no confusion can arise.} that corresponds to the quantum observable $X_A$ in (\ref{eq:35a})-(\ref{eq:35}), the meaning of the other variables is obvious, $a,b,c,d$ and $e$ are shorthand notations to denote the settings and additional variables of any apparata located close to $A, B, C,D$ and $E$, respectively, and $\lambda$ are additional (local) parameters. As previouly stipulated, $x_A$ (as well as any other dichotomic variable pertaining to $A, B, C$ and $D$) does not depend on the settings $e$ of any apparata or variables located close to $E$, and $x_E$ (as well as any other dichotomic variable pertaining to $E$) does not depend on the settings $a,b,c,d$ of any apparata or variables located close to $A, B, C$ or $D$.

Let now Esther measure one of her dichotomic variables $x_E, y_E$ or $z_E$ and get one of the following outcomes
\beq
\label{eq:resE}
 x_E = \gamma, \quad  y_E = \delta, \quad z_E = \epsilon \quad
(\gamma,  \delta,  \epsilon = \pm 1).
\eeq
By usual local-realistic arguments, any of the possible outcomes (\ref{eq:resE}) must be pre-determined and exist independently of observation.
Assume now that the quantum mechanical predictions are valid, in the sense that the correlations 
 (\ref{eq:35a})-(\ref{eq:35}) correctly reproduce the measurements of $A, B, C, D$ and $E$.
It is then straightforward to see that Eqs.\
 (\ref{eq:35a})-(\ref{eq:35}) translate into (we suppress for conciseness the dependence of all the variables on $a,b,c,d$ and $\lambda$)
\barr
\label{eq:35eqsA}
& & z_A x_B x_C  = 1, \\
& & x_A z_B z_D  = 1, \\
& & y_A y_B  = \epsilon, \label{eq:yy} \\
& & y_A z_C y_D  = 1, \\
& & x_A y_C  = \delta,  \label{eq:xy} \\
& & z_A x_D  = \gamma,  \label{eq:zx} \\
& & y_B y_C  x_D  = -1, \label{eq:yyx} \\
& & z_B z_C  = \gamma,  \label{eq:zz} \\
& & x_B y_D  = -\delta,  \label{eq:xyD} \\
& & x_C z_D  = \epsilon, \label{eq:xz} \\ 
& & x_A x_B z_C x_D = -\epsilon, \\
& & x_A y_B x_C y_D = -\gamma, \\
& & y_A x_B y_C z_D = \gamma, \\
& & y_A z_B x_C x_D = \delta, \\
& & z_A y_B z_C z_D = \delta, \\
& & z_A z_B y_C y_D = \epsilon. 
\label{eq:35eqs}
\earr
Simple scrutiny of Eqs.\ (\ref{eq:35eqsA})-(\ref{eq:35eqs})
shows that they admit no solution for any of the 8 possible combinations (\ref{eq:resE}). A nonlocal hidden variable model of the type 
(\ref{eq:nonloc})-(\ref{eq:vardef}) is therefore unable to reproduce the quantum correlations  (\ref{eq:35a})-(\ref{eq:35}).
This is what we wanted to prove.

An experimental test of the nonlocal hidden-variable theories considered in this Letter involves some conceptual subtleties that are worth discussing. At the same time, this discussion will help us sharpen our argument. We start by observing that most of the sixteen measurements (\ref{eq:35eqsA})-(\ref{eq:35eqs}) are incompatible [not all: simultaneous measurements of (\ref{eq:yy}) and (\ref{eq:xz}), (\ref{eq:xy}) and (\ref{eq:xyD}), (\ref{eq:zx}) and (\ref{eq:zz}) are possible], so $A,B,C$ and $D$ must repeat their experiment at least $16-3=13$ times, each time with different settings of the polarizers, in accord with the left hand sides of Eqs.\ (\ref{eq:35eqsA})-(\ref{eq:35eqs}). 

A possible experiment is the following. $A,B,C$ and $D$, space-like separated from each other (and from $E$), start by sharing $N(\gg 13)$ identical ensembles of five particles in state (\ref{eq:psi5}). Then, in each run, they independently set their polarizers (their decisions being space-like separated) and measure their observables.
Having completed their $N$ measurements \footnote{There are $4^4=256$ possible distinct measurements. Notice that the \emph{whole} set of $N$ measurements of each party can be required to be at a space-like distance from the set of $N$ measurements of any other party.}, they communicate to each other their polarizer settings and check which subset of joint measurements coincides with the left hand sides of Eqs.\ (\ref{eq:35eqsA})-(\ref{eq:35eqs}); at the same time, they share their outcomes and multiply those relative to the ``useful" measurement subset, obtaining a list of digits ($\pm1$) 
\footnote{Although this is immaterial for our argument, we observe that, at this stage, they do not know whether these digits are the same as those obtained by $E$ and given by the right hand sides of Eqs.\ (\ref{eq:35eqsA})-(\ref{eq:35eqs}): if they want to check, they must wait for $E$ to communicate her outcomes by local (classical) means.}. Finally, they endeavor to give a nonlocal explanation of their outcomes: for example, if, in a given experimental run, the joint observable (\ref{eq:yyx}) was measured, Bob (after having compared his outcomes with those of Charlie and Diana) will be allowed to justify his finding by (nonlocally) concluding that 
\beq
y_B(a,b,c,d,\lambda)  = -y_C(a,b,c,d,\lambda)  x_D(a,b,c,d,\lambda). \label{eq:yyxBis} \\
\eeq
By assuming that $E$ is independent [in the sense of (\ref{eq:nonloc})-(\ref{eq:vardef})] and of course at a space-like distance, the products of measurement outcomes (Einstein's ``elements of reality" \cite{EPR} \footnote{Interestingly, these elements of reality pertain to products of measurement outcomes and not to single measurement outcomes.}) 
obtained by $A,B,C$ and $D$ in their measurements must have been determined in advance (before the measurements were actually perfomed). 
By assuming that quantum mechanics is valid and experimental uncertainties/errors negligible, these products of outcomes must coincide with the right-hand sides of Eqs.\ (\ref{eq:35eqsA})-(\ref{eq:35eqs}). However, the system (\ref{eq:35eqsA})-(\ref{eq:35eqs}) admits no solution. By further assuming that the state of the particles yields a complete description of the latter and is the same in all experiments (a form of realism), the incompatibility between quantum mechanics and nonlocal hidden-variable theories is estabilished. It goes without saying that the above argument hinges on some form of counterfactual definiteness \cite{counterfactual}.

We close with a few comments. First of all, we notice that Esther plays no special role. 
Similar results are obtained by permuting the parties $A,B,C,D$ and $E$ (e.g., if one assumes that the outcomes of $B, C, D$ and $E$ nonlocally depend on each other but do not depend on those of $A$, and so on). 
Second, the situation depicted in Fig.\ \ref{fig:abcde} is instrumental in making our argument, the only relevant assumptions being 
(\ref{eq:nonloc})-(\ref{eq:vardef}).
Third, it may appear artificial to set one party aside from the other four. However, as observed in the paragraph that precedes Eq.\ (\ref{eq:nonloc}), a fully nonlocal hidden-variable model cannot be discriminated from quantum mechanics and invoking the lack of symmetry among the five parties as a possible way out of the nonlocality-vs-quantum-mechanics puzzle seems to us even more artificial.
Finally, it is natural to ask whether the present analysis can be extended to a larger number of parties. An extension to six parties is certainly possible, as perfect MMES are known to exist for $N=6$ qubits \cite{MMES}. An extension to $N \geq 7$ (and possibly $N\to \infty$) is nontrivial and under investigation. 

We have proved that nonlocal hidden-variable models satisfying conditions (\ref{eq:nonloc})-(\ref{eq:vardef}) are in contradiction with quantum mechanics.
The results obtained in this Letter suggest that in the incompatibility between quantum mechanics and local realism established by Bell, the real problem is realism, namely the capability of defining a reality that is independent of measurement.

\acknowledgements
I would like to thank P. Facchi, G. Florio, G. Parisi and A. Scardicchio for interesting conversations on multipartite entanglement.


\end{document}